\documentclass{article}
\usepackage{spconf}
\usepackage{microtype}
\usepackage{amsmath,amssymb,amsfonts,mathtools}
\usepackage{booktabs}
\usepackage{multirow}
\usepackage{tabularx}
\usepackage{array}
\usepackage{float}
\usepackage{adjustbox}
\usepackage{xcolor}
\usepackage{url}
\usepackage[numbers,sort&compress]{natbib}
\usepackage[hidelinks]{hyperref}
\newcolumntype{C}{>{\centering\arraybackslash}X}
\graphicspath{{./}{figures/}}
\title{Flow-HOA: Generative Joint Optimization\\for Ambisonics Encoding via Flow Matching}
\name{\small Yuhuan You\textsuperscript{1}, Yufan Qian\textsuperscript{1}, Tianshu Qu\textsuperscript{1}, Bin Wang\textsuperscript{2}, Xueyang Lv\textsuperscript{3}}
\address{$^{1}$State Key Laboratory of General Artificial Intelligence,\\
School of Intelligence Science and Technology,\\
Peking University, Beijing, China\\
$^{2}$Beijing Xiaomi Mobile Software Co., Ltd\\
$^{3}$Xiaomi Communications Co., Ltd}

\begin{document}

\maketitle

\begin{center}
\scriptsize\itshape
Accepted for presentation at\\
AES Europe 2026 Convention\\
(AES 160th Convention)\\
Copenhagen, Denmark, May 28-30, 2026.\\
This is the authors' version prepared\\
without AES graphical style or template.
\end{center}

\begin{abstract}
Higher-Order Ambisonics (HOA) encoding from sparse, irregular microphone arrays remains a critical challenge for consumer spatial audio capture in immersive communication and XR. We propose Flow-HOA, a generative framework that jointly optimizes a multi-dimensional objective encompassing time-domain, spectral, and spatial fidelity while producing a deployable, time-invariant bank of Finite Impulse Response (FIR) encoding filters. Using conditional flow matching, the model learns to map a simple prior distribution to the target distribution of FIR filter coefficients. Training is guided by a composite loss that balances time-domain waveform fidelity, multi-resolution spectral consistency, sub-band energy preservation, and spatial directivity constraints. Objective evaluations on synthetically simulated data demonstrate improved performance over strong model-based baselines in both signal fidelity and spatial accuracy metrics. Subjective listening tests on real microphone array recordings further confirm that Flow-HOA yields higher overall sound quality with reduced artifacts, demonstrating generalization from synthetic training data to real-world capture conditions.
\end{abstract}

\section{Introduction}
\label{sec:intro}

The proliferation of virtual reality (VR) and augmented reality (AR) devices has established spatial audio capture and rendering as a pivotal technology for enhancing user immersion. Higher-Order Ambisonics (HOA), a powerful and loudspeaker-independent format, represents the 3D sound field using a basis of spherical harmonic functions and has become a cornerstone technology in this domain \cite{Zotter2019Ambisonics}. Its core advantage lies in the ability to smoothly rotate the entire sound field, which is crucial for maintaining spatial consistency in head-tracked VR/AR applications.

In theory, specially designed spherical microphone arrays allow for the direct computation of HOA signals via an analytical plane-wave decomposition \cite{Meyer2002}. However, this ideal scenario is often impractical for consumer-grade devices like smartphones and wearable glasses, which are typically constrained by industrial design to feature only sparse, irregularly distributed, and asymmetric microphone arrays. From a signal processing perspective, estimating HOA coefficients from such arbitrary geometries constitutes an ill-posed inverse problem, particularly in the high-frequency regime where the wavelength is comparable to the inter-microphone spacing. This leads to severe spatial aliasing and ambiguity, making robust encoding a formidable challenge.

To accommodate these arbitrary geometries, researchers have developed various methods \cite{gayer2025ambisonicsencoderwearablearray, Zaunschirm2018, Berebi_2025, you2025, Benesty2008, Perotin2019}, broadly categorized into signal-independent and signal-dependent approaches \cite{rafaely2015Fundamentals}. Signal-independent methods aim to design a universal, fixed filter bank derived solely from the array geometry. Prominent baselines include Ambisonics Signal Matching (ASM) \cite{gayer2025ambisonicsencoderwearablearray} and regularized least-squares variants \cite{Zaunschirm2018, you2025}. While computationally efficient, these model-based methods face a fundamental trade-off between spatial directivity and white noise gain (WNG) \cite{Poletti2005}. Heavy regularization required to stabilize the filters often results in temporal smearing, spectral coloration, and a loss of spatial detail, failing to preserve the perceptual fidelity of the sound scene.

In contrast, signal-dependent methods adapt their processing to the instantaneous content of the sound field. One branch, parametric synthesis, estimates source parameters like direction-of-arrival (DOA) to synthesize virtual HOA signals. However, its accuracy is fundamentally bottlenecked by the difficulty of robust multi-source localization with sparse arrays, often leading to unstable spatial images and "musical noise" artifacts \cite{Benesty2008}. Another branch, end-to-end neural mapping \cite{Perotin2019}, learns a direct non-linear transformation from microphone signals to HOA coefficients. While promising, these deep learning methods typically operate as "black boxes" with significant computational overhead and high latency, rendering them unsuitable for real-time deployment on battery-powered mobile devices.

A deeper analysis reveals that the limitations of prior methods largely stem from their reliance on simplified optimization objectives. Whether matching ideal plane wave coefficients analytically or minimizing a simple mean squared error (MSE), these objectives fail to capture the complexities of human auditory perception in dynamic acoustic scenes. To overcome this bottleneck, we argue for a shift towards a data-driven optimization paradigm based on a composite multi-domain metric. However, such a comprehensive objective is highly non-convex and intractable for traditional convex optimization solvers.

To this end, we propose {Flow-HOA}, a novel framework that reformulates HOA filter design as a generative joint optimization task. By leveraging Conditional Flow Matching (CFM) \cite{lipman2023flowmatchinggenerativemodeling}, we learn a continuous probability path that transforms a physics-informed prior into an optimal filter. This approach allows us to navigate the complex optimization landscape effectively, combining the robustness of physical priors with the multi-domain fidelity of data-driven optimization. Our core contributions are as follows:

\textbf{A generative joint optimization framework.} We depart from traditional analytical derivations, instead formulating HOA filter design as a generative modeling task. This allows the solver to escape local minima inherent in analytical regularizations.

\textbf{A composite multi-domain optimization objective.} We introduce a multi-dimensional loss function that strictly constrains the filter across time, frequency, and spatial domains. This ensures that the generated filters preserve waveform phase, spectral timbre, and directional energy distribution simultaneously.

\textbf{Efficient, deployment-ready FIR filters.} A key distinction of our work is that the heavy neural computation is restricted to the offline design phase. The final output is a set of standard Finite Impulse Response (FIR) filters, ensuring low-latency and low-power execution on existing consumer hardware.

The rest of this paper is organized as follows. Section \ref{sec:methods} details the proposed Flow-HOA method. Section \ref{sec:exp} presents the experimental setup, objective and subjective evaluations. Finally, Section \ref{sec:conclusion} provides concluding remarks.

\section{Proposed Methods}
\label{sec:methods}

This section details the Flow-HOA framework. We bridge the gap between simplified analytical models and the complexities of practical audio reality by formulating the filter design as a generative process. Our approach comprises three stages: constructing a physics-informed prior to ground the solution, defining a composite multi-domain objective to jointly optimize time, frequency, and spatial fidelity, and employing conditional flow matching to robustly synthesize the final Finite Impulse Response (FIR) filters. Since we train an independent generator for each HOA channel, the framework scales to arbitrary Ambisonics orders by simply adding more channel-specific models.

\subsection{Physics-Informed Prior Filter Design}

Ambisonics provides a complete, hierarchical representation of a three-dimensional sound field using a basis of Spherical Harmonics (SH) \cite{Zotter2019Ambisonics}. For a plane wave signal $s(t)$ arriving from a direction $\Omega_k = (\theta_k, \phi_k)$, its ideal HOA signal $b_{nm}^\sigma(t)$ is defined as:
\begin{equation}
b_{nm}^{\sigma}(t) = s(t) \cdot Y_{nm}^{\sigma}(\theta_k,\phi_k)
\end{equation}
where $Y_{nm}^\sigma$ is the real-valued SH of order $(n,m)$ and type $\sigma \in \{-1, +1\}$ (denoting the sine or cosine component, respectively). In this work, we define the task of HOA encoding as designing a Finite Impulse Response (FIR) filter matrix $\mathbf{H} \in \mathbb{R}^{C \times Q \times L}$, where $C=(N+1)^2$ is the number of HOA channels for a target order $N$, $Q$ is the number of microphones, and $L$ is the FIR filter length. This matrix accurately estimates the $C$ ideal HOA signals from the signals $\mathbf{x}(t)$ captured by $Q$ microphones.

\begin{figure}[t]
\centering
\includegraphics[width=0.96\linewidth]{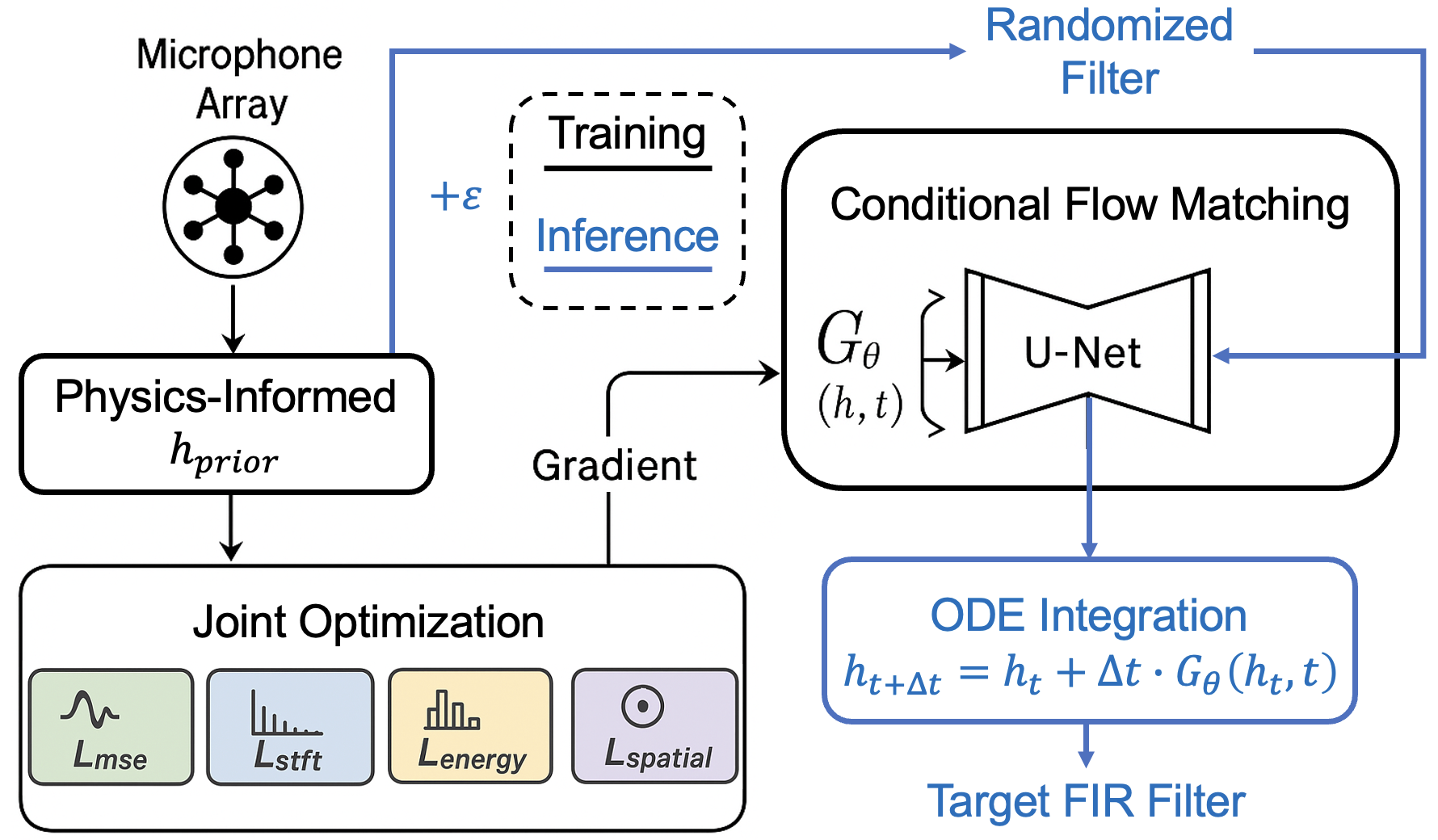}
\caption{Overview of Flow-HOA: The training phase learns a vector field from joint loss gradients; the inference phase integrates this field to refine a physics-informed prior into optimal FIR filters.}
\label{fg:frame}
\end{figure}

The initialization of our generative process is critical. Instead of starting from random noise, which requires the model to learn basic wave propagation physics from scratch, we construct a {physics-informed prior filter}, $\mathbf{h}_{\text{prior}}$ (see ``Physics-Informed $\mathbf{h}_{\text{prior}}$'' in Fig.~\ref{fg:frame}). This is achieved by solving a constrained time-domain least-squares problem over all $K$ measured spatial directions that incorporates the measured impulse responses ($\mathbf{d}_{k,q}$) of the specific array. This approach, grounded in robust broadband array processing \cite{Poletti2005}, aims to ensure that for a unit impulse input from any direction $\Omega_k$, the system's output approximates an ideal, delayed delta function weighted by the corresponding SH value:
\begin{equation}
J(\mathbf{H}) = \sum_{k=1}^{K} \left\| \sum_{q=1}^{Q} (\mathbf{h}_{q} * \mathbf{d}_{k,q}) - \mathbf{y}_{k} \right\|_2^2 + \gamma \|\mathbf{H}\|_F^2
\end{equation}
where $K$ is the total number of measured spatial directions, $\mathbf{h}_q \in \mathbb{R}^L$ is the FIR filter for the $q$-th microphone, $\mathbf{d}_{k,q} \in \mathbb{R}^{L_{IR}}$ is the measured impulse response from direction $\Omega_k$ to microphone $q$, $\mathbf{y}_{k}(n) = Y_{nm}^{\sigma}(\Omega_k)\delta[n-n_0]$ is the target impulse response with modeling delay $n_0$, and $\gamma$ is a Tikhonov regularization parameter. Casting this as a matrix equation $\mathbf{D}\mathbf{h} = \mathbf{y}$, the closed-form solution is obtained via the Moore-Penrose pseudoinverse \cite{Golub2013Matrix}: $\mathbf{h}_{\text{prior}} = (\mathbf{D}^{\top}\mathbf{D} + \gamma\mathbf{I})^{-1}\mathbf{D}^{\top}\mathbf{y}$.

\textbf{Rationale for Refinement:} While this analytical prior $\mathbf{h}_{\text{prior}}$ provides a solid starting point, it is fundamentally limited by the "spatial aliasing" phenomenon inherent in sparse arrays. At high frequencies, the least-squares solution must aggressively trade off between directivity and white noise gain, often resulting in severe spectral coloration and spatial blurring artifacts. The subsequent neural stages of our framework are explicitly designed to correct these physical limitations by "relaxing" the strict analytical constraints in favor of perceptual plausibility.

\subsection{Joint Optimization Objective}

To transcend the limitations of impulse-matching, we employ a data-driven joint objective, $\mathcal{L}_{\text{joint}}$, evaluated on continuous audio signals. Here, the ideal signal $\mathbf{y}_{\text{ideal}}$ for a given source $s(t)$ from direction $\Omega_k$ is obtained analytically as $\mathbf{y}_{\text{ideal}}(t) = s(t) \cdot Y_{nm}^{\sigma}(\Omega_k)$ (cf.\ Eq.~1). The optimal filter $\mathbf{h}^*$ is defined as the minimizer of the expected loss over a distribution of natural acoustic scenes $\mathcal{S}$:
\begin{equation}
\mathbf{h}^* = \arg\min_{\mathbf{h}} \mathbb{E}_{s \sim \mathcal{S}} [\mathcal{L}_{\text{joint}}(\mathbf{h}; s)]
\end{equation}
The loss function is a weighted sum of four components, each targeting a specific fidelity dimension:
\begin{equation}
\begin{aligned}
\mathcal{L}_{\text{joint}} =\;&
\lambda_{\text{mse}} \mathcal{L}_{\text{mse}}
+ \lambda_{\text{stft}} \mathcal{L}_{\text{stft}} \\
&+ \lambda_{\text{energy}} \mathcal{L}_{\text{energy}}
+ \lambda_{\text{spatial}} \mathcal{L}_{\text{spatial}}
\end{aligned}
\end{equation}

\textbf{Time-Domain Fidelity ($\mathcal{L}_{\text{mse}}$):} To ensure fundamental waveform consistency and phase alignment, we include a Mean Squared Error (MSE) term. It penalizes the sample-level discrepancy between the estimated signal $\mathbf{z}_{\text{est}}$ and the ideal signal $\mathbf{y}_{\text{ideal}}$:
\begin{equation}
\mathcal{L}_{\text{mse}} = \frac{1}{B N'} \sum_{b=1}^{B} \sum_{n=1}^{N'} (\mathbf{z}_{\text{est}, c}^{(b)}[n] - \mathbf{y}_{\text{ideal}, c}^{(b)}[n])^2
\end{equation}
While MSE is crucial for phase correctness, relying on it alone often leads to muddy, over-smoothed high-frequency responses due to its sensitivity to microscopic time shifts.

\textbf{Spectral \& Timbral Consistency ($\mathcal{L}_{\text{stft}}$):} To address the "spectral coloration" artifacts common in beamforming, we introduce a multi-resolution Short-Time Fourier Transform (STFT) loss \cite{Kong2020HiFiGAN}. By evaluating spectral magnitude and log-magnitude differences across $M$ resolutions, this term forces the filter to maintain a natural timbre and prevents the "metallic" sound associated with comb-filtering effects:
\begin{equation}
\begin{split}
\mathcal{L}_{\text{stft}} = & \sum_{i=1}^{M} \left( \frac{1}{B} \sum_{b=1}^{B} \frac{\| |Y_c^{(b, i)}| - |Z_c^{(b, i)}| \|_F}{\| |Y_c^{(b, i)}| \|_F + \epsilon} \right. \\
& \left. + \frac{1}{B T_i F_i} \sum_{b=1}^{B} \sum_{t,f} \left| \log(|Y_c^{(b, i)}|) - \log(|Z_c^{(b, i)}|) \right| \right)
\end{split}
\end{equation}

\textbf{Sub-band Energy Preservation ($\mathcal{L}_{\text{energy}}$):} Inspired by the sub-band decomposition used in perceptual audio coding \cite{Painter2000}, we propose a loss to prevent the optimizer from collapsing the energy in difficult frequency bands (e.g., above the spatial aliasing frequency). We enforce energy conservation across $P$ triangular frequency bands:
\begin{equation}
\mathcal{L}_{\text{energy}} = \frac{1}{B P} \sum_{b=1}^{B} \sum_{j=1}^{P} (\log E_{\text{est}}^{(b)}(j) - \log E_{\text{ideal}}^{(b)}(j))^2
\end{equation}

\textbf{Spatial Directivity ($\mathcal{L}_{\text{spatial}}$):} Finally, we propose a spatial fidelity loss that explicitly constrains the spatial distribution of energy. Since perfect waveform matching is impossible above the aliasing frequency, this term relaxes the phase constraint and focuses on ensuring that the {directionality} of the sound field is correct. We perform virtual beamforming towards $K'$ directions and minimize the log-energy difference:
\begin{equation}
\mathcal{L}_{\text{spatial}} = \frac{1}{B K'} \sum_{b=1}^{B} \sum_{k=1}^{K'} (\log E_{\text{est}}^{(b)}(\Omega_k) - \log E_{\text{ideal}}^{(b)}(\Omega_k))^2
\end{equation}

\subsection{Generative Synthesis via Flow Matching}

Directly optimizing the high-dimensional, non-convex joint loss function $\mathcal{L}_{\text{joint}}$ is notoriously unstable and prone to local minima. To robustly solve for optimal filters, we re-frame the problem as learning a vector field that transports the prior distribution to the optimal posterior. We adopt the \textbf{Conditional Flow Matching (CFM)} paradigm.

The core idea is to train a neural network $G_{\theta_c}$ (parameterized as a 1D U-Net) to approximate the {smoothed gradient vector field} of the objective function. Effectively, the network acts as a "learned optimizer" or "meta-learner." For a given filter $\mathbf{h}$, the network predicts the optimal update direction $\mathbf{g}_{c}(\mathbf{h}) \approx -\nabla_{\mathbf{h}} \mathcal{L}_{\text{joint}}$.

\textbf{Training Strategy:} In each iteration, we sample a perturbed filter $\mathbf{h}_{\text{rand}}$ near the prior $\mathbf{h}_{\text{prior}}$. We then compute the {true} gradient of the composite loss $\mathcal{L}_{\text{joint}}$ via automatic differentiation through the physical simulation of the array. This noisy ground-truth gradient is used to supervise the generator $G_{\theta_c}$. The objective is to minimize the flow matching loss $\mathcal{L}_{\text{FM}}$, defined as the mean squared error between the network prediction and the target gradient field:
\begin{equation}
\mathcal{L}_{\text{FM}}(\theta_c) = \mathbb{E}_{\mathbf{h}, t, s} \left[ \left\| G_{\theta_c}(\mathbf{h}, t) - \bar{\mathbf{g}}_{c}(\mathbf{h}) \right\|_2^2 \right]
\end{equation}
where $\bar{\mathbf{g}}_{c}$ is the Exponential Moving Average (EMA) of the raw gradients, which stabilizes the training dynamics.

\textbf{Inference as ODE Integration:} Once trained, $G_{\theta_c}$ defines a continuous vector field governing the evolution of the filter coefficients. Theoretically, this corresponds to an Ordinary Differential Equation (ODE):
\begin{equation}
\frac{d\mathbf{h}_t}{dt} = G_{\theta_c}(\mathbf{h}_t, t)
\end{equation}
Note that the flow time $t \in [0, 1]$ is a parameter of the generative process and is unrelated to the audio sample rate. During inference, we bypass the heavy computation of gradients. Starting from the physics-informed prior $\mathbf{h}_{\text{prior}}$, we numerically integrate this ODE using the forward Euler method with $N_{\text{steps}}=100$ uniformly spaced steps and a fixed step size $\Delta t = 1/N_{\text{steps}}$:
\begin{equation}
\mathbf{h}_{t+\Delta t} = \mathbf{h}_t + \Delta t \cdot G_{\theta_c}(\mathbf{h}_t, t)
\end{equation}
This process ``flows'' the initial analytical solution along the learned manifold, gradually refining it to satisfy the complex multi-domain constraints, resulting in the final optimized filter $\mathbf{h}_{1}$.

\subsubsection{Network Architecture and Training Stability.}
The generator network $G_{\theta_c}$ is implemented as a 1D U-Net architecture tailored for filter generation. It features a symmetric encoder-decoder structure with skip connections to preserve high-frequency details. Unlike standard convolutional networks, we utilize Group Normalization (configured with a single group to emulate Layer Normalization) followed by SiLU (Sigmoid Linear Unit) activations throughout the network to enhance generation quality and training stability. The time step $t$ is encoded using sinusoidal position embeddings and projected via dense layers to modulate the feature maps at every resolution level. To ensure robust convergence against the high variance of physical gradients, we employ three specific stabilization strategies during training:
(1) \textit{Noise Injection}: We perturb the prior filter $\mathbf{h}_{\text{prior}}$ with Gaussian noise ($\sigma=0.01$) to improve the vector field's robustness around the manifold;
(2) \textit{Gradient Smoothing}: Instead of fitting the instantaneous noisy gradient, the network targets an Exponential Moving Average (EMA) of the gradients with a decay factor $\beta=0.9$, effectively filtering out high-frequency optimization noise;
(3) \textit{Target Clipping}: The target gradient values are explicitly clipped to the range $[-1, 1]$ to prevent gradient explosion during the initial training phase.

\section{Experiments}
\label{sec:exp}

\begin{figure}[t]
\centering
\includegraphics[width=0.88\linewidth]{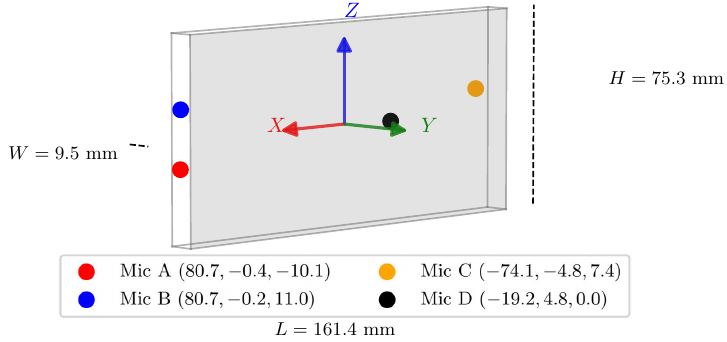} 
\caption{Microphone array configuration of the SPMA.}
\label{fg:arraypos}
\end{figure}

\subsection{Experimental Setup}

\subsubsection{Hardware and Physical Measurements}
\begin{figure}[t]
\centering
\includegraphics[width=0.92\linewidth]{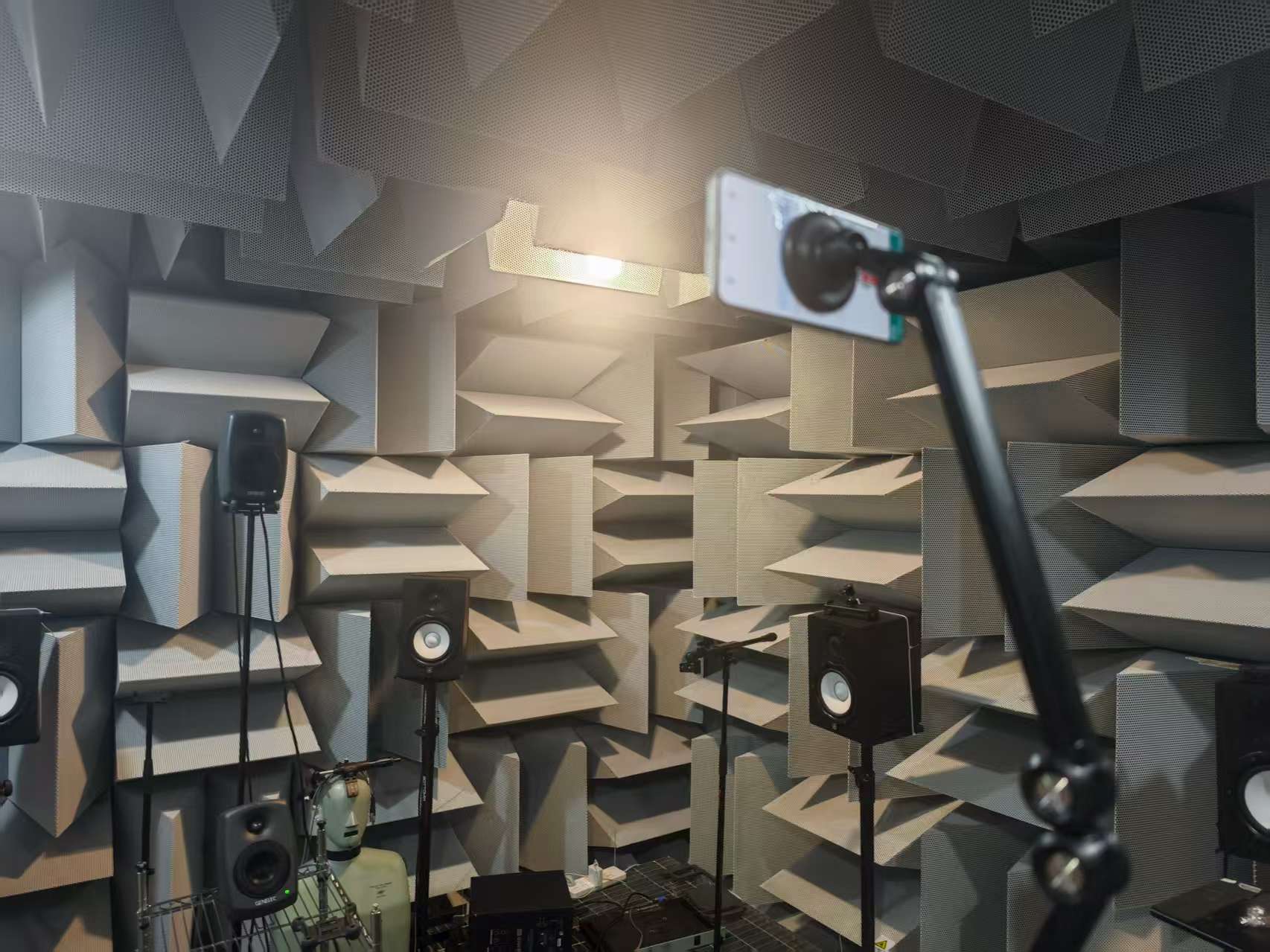}
\caption{Anechoic-chamber measurement setup used to capture the Array Impulse Responses (AIRs).}
\label{fg:chamber}
\end{figure}

The target device for this study is a realistic smartphone microphone array (SPMA) prototype. It features $Q=4$ MEMS microphones arranged in an asymmetric configuration, as illustrated in Figure~\ref{fg:arraypos}. All audio is processed at a sampling rate of 48\,kHz, and the FIR filter length is set to $L=1024$ samples. The asymmetry is intentionally designed to mitigate spatial aliasing ambiguities common in symmetric uniform arrays. 
To accurately capture the array's physical properties (shadowing, diffraction, and scattering effects), we conducted extensive physical measurements in a fully anechoic chamber (Figure~\ref{fg:chamber}). The Impulse Responses (IRs) were measured from $K=180$ discrete spatial directions. These directions encompass 36 azimuth angles (at $10^\circ$ intervals) across 5 elevation bands $(0^\circ, \pm 30^\circ, \pm 60^\circ)$. A Genelec loudspeaker was excited by a logarithmic Exponential Sine Sweep (ESS) signal \cite{farina2007advancements} to retrieve the impulse responses for each direction. These measured IRs constitute the {ground-truth array manifold} used for both the physics-informed prior calculation and the training data synthesis.

\subsubsection{Dataset and Dynamic Simulation}
To ensure the model generalizes to diverse acoustic contents, we utilized the FSD50K dataset \cite{fonseca2022fsd50k} as the source of dry audio events. Unlike traditional approaches that rely on pre-generated datasets, we implemented an \textit{on-the-fly dynamic mixing pipeline}. During each training iteration, a dry source clip is randomly sampled and convolved with the measured Array Impulse Responses (AIRs) corresponding to a randomly selected direction $\Omega_k$ from the $K=180$ discrete measurements. This strategy effectively acts as a continuous data augmentation mechanism, preventing the network from overfitting to specific source-direction pairings.
Simultaneously, the ground-truth target signals (ideal HOA) are generated analytically as $\mathbf{y}_{\text{ideal}}(t) = s(t) \cdot Y_{nm}^{\sigma}(\Omega_k)$, as defined in Eq.~(1). This ensures that the training targets are mathematically precise and free from measurement noise or aliasing artifacts inherent in the microphone signals.

\subsubsection{Baselines and Implementation Details}
We compare Flow-HOA against the industry-standard Ambisonics Signal Matching (ASM) method. The ASM baseline computes the encoding matrix analytically in the Short-Time Fourier Transform (STFT) domain using regularized least-squares inversion, serving as a representative benchmark for linear, time-invariant (LTI) spatial filters.

For the proposed Flow-HOA, the framework was implemented in PyTorch. We trained an independent 1D U-Net generator for each of the $C=25$ HOA channels (4th order). The training process was conducted for 50 epochs using the AdamW optimizer \cite{Loshchilov2019AdamW}. During inference, we employ a first-order Euler ODE solver to numerically integrate the learned vector field, transporting the physics-informed prior to the optimal posterior.
The complete set of hyperparameters and loss function weights is detailed in Table \ref{tab:hyperparams}. Notably, a high weight is assigned to $\lambda_{\text{mse}}$ to prioritize phase alignment, while spatial and spectral losses act as regularizers to improve overall fidelity.

\begin{table}[t]
\centering
\small
\caption{Implementation Hyperparameters and Loss Configurations.}
\label{tab:hyperparams}
\begin{adjustbox}{max width=\columnwidth}
\begin{tabular}{l c}
\toprule
\textbf{Parameter} & \textbf{Value} \\
\midrule
\multicolumn{2}{l}{\textit{Training Setup}} \\
Optimizer & AdamW \\
Learning Rate & $1 \times 10^{-5}$ \\
Batch Size & 256 \\
Filter Length ($L$) & 1024 \\
Inference Steps ($N_{\text{steps}}$) & 100 \\
\midrule
\multicolumn{2}{l}{\textit{Loss Weights} ($\mathcal{L}_{\text{joint}}$)} \\
$\lambda_{\text{mse}}$ & 50.0 \\
$\lambda_{\text{stft}}$ & 0.1 \\
$\lambda_{\text{energy}}$ & 0.1 \\
$\lambda_{\text{spatial}}$ & 0.1 \\
\bottomrule
\end{tabular}
\end{adjustbox}
\end{table}

\subsection{Evaluation Metrics}

We define a comprehensive suite of objective metrics to evaluate the synthesized HOA signals from two critical perspectives: {Signal Fidelity} (timbre and phase) and {Spatial Fidelity} (localization and sound stage). All evaluations are performed on the unseen test set.

\subsubsection{Signal Fidelity}
First, we assess the quality of the reconstructed waveform, which directly impacts the perceived transparency and coloration of the sound.

\textbf{1) Scale-Invariant Signal-to-Distortion Ratio (SI-SDR) [dB]:} We use SI-SDR \cite{leroux2019sdr} to measure time-domain waveform accuracy. Unlike standard SNR, SI-SDR is invariant to global gain scaling and is highly sensitive to phase distortions. A higher SI-SDR indicates that the transient structure and phase coherence of the source signal are well-preserved.

\textbf{2) Log-Spectral Distance (LSD):} To evaluate frequency-domain fidelity, we compute the distance between the log-magnitude spectrograms of the estimated and ideal signals \cite{gray1976distance}. A lower LSD score signifies less spectral coloration, implying that the filter bank has successfully avoided the comb-filtering artifacts often associated with aggressive beamforming.

\subsubsection{Spatial Fidelity}
Second, we evaluate the structural integrity of the reconstructed 3D sound field, which determines the immersion in VR/AR environments.

\textbf{1) Spatial Power Map divergence (SPM-KL):} Following the use of KL divergence on spatial power maps for Ambisonics evaluation \cite{Qiao2025}, we quantify the accuracy of the spatial energy distribution using the Kullback-Leibler (KL) divergence \cite{Kullback1951}. A value of zero indicates perfect spatial reconstruction, with higher values indicating greater divergence. We project both the estimated and ideal HOA signals onto a spherical grid of $N_g$ points via plane-wave beamforming \cite{rafaely2015Fundamentals} to obtain their Spatial Power Maps ($P_{\text{est}}$ and $P_{\text{ideal}}$).
\begin{equation}
D_{KL}(P_{\text{ideal}} \| P_{\text{est}}) = \sum_{i=1}^{N_g} P_{\text{ideal}}(i) \log \left(\frac{P_{\text{ideal}}(i)}{P_{\text{est}}(i)+\epsilon}\right)
\end{equation}
This probabilistic metric goes beyond simple peak-matching; it penalizes spatial blurring (widening of the source) and spurious ghost sources (sidelobes).
    
\textbf{2) Directional Gain Consistency (DGC) [dB]:} A fundamental design criterion for Ambisonics systems is that the encoding gain should remain constant for all source directions \cite{Zotter2019Ambisonics, Heller2008}. Motivated by this principle, we propose the DGC metric to quantify response uniformity. DGC measures the standard deviation of the energy gain (in dB) across all $K$ test directions:
\begin{equation}
G_k = 10 \log_{10} \left( \frac{E(\mathbf{b}_{\text{est}, k})}{E(\mathbf{x}_k)} \right),\mathrm{DGC}=\operatorname{std}\left(G_1, \ldots, G_K\right)
\end{equation}
A lower DGC value indicates a more isotropic response, ensuring that sources do not unnaturally fluctuate in volume as they move around the listener.

\subsection{Objective Evaluation}

The objective evaluation was strictly performed on the unseen FSD50K evaluation set. To ensure statistical significance, we randomly selected 100 audio clips and, for each clip, simulated the source at 50 randomly sampled spatial directions. This resulted in a total of 5000 unique evaluation instances, covering a diverse range of spectral contents and spatial locations. The quantitative results are summarized in Table~\ref{tab:results}.

\begin{table}[t]
\centering
\small
\caption{Objective Evaluation Results on the Unseen Test Set.}
\label{tab:results}
\setlength{\tabcolsep}{4pt}
\begin{adjustbox}{max width=\columnwidth}
\begin{tabularx}{\columnwidth}{lCCCC}
\toprule
\textbf{Method} & \textbf{SI-SDR [dB] $\uparrow$} & \textbf{LSD $\downarrow$} & \textbf{SPM-KL $\downarrow$} & \textbf{DGC [dB] $\downarrow$} \\
\midrule
ASM & -13.72 & 11.12 & 1.44 & 2.17 \\
Flow-HOA & \textbf{-7.31} & \textbf{5.07} & \textbf{1.14} & \textbf{0.84} \\
\bottomrule
\end{tabularx}
\end{adjustbox}
\end{table}

As evidenced by Table~\ref{tab:results}, the proposed Flow-HOA framework significantly outperforms the conventional ASM baseline across all four metrics. We attribute these improvements to the generative model's ability to navigate the trade-off between noise robustness and signal fidelity more effectively than linear analytical methods.

\subsubsection{Analysis of Signal Fidelity}
In the time domain, Flow-HOA achieves a substantial {6.41 dB improvement in SI-SDR}. Analytical methods like ASM rely on Tikhonov regularization to prevent filter explosion at low frequencies and near spatial aliasing points. While this stabilizes the filter, it inevitably introduces phase distortion and "temporal smearing," destroying the fine structure of transient signals. Flow-HOA, guided by the $\mathcal{L}_{\text{mse}}$ and physics-informed prior, learns to preserve phase linearity, resulting in a cleaner waveform reconstruction.
In the frequency domain, the {LSD score is reduced by over 50\%} (from 11.12 to 5.07). High LSD in the baseline is indicative of "spectral coloration"—unnatural peaks and dips in the frequency response caused by the constructive and destructive interference of the beamformer sidelobes. By optimizing the $\mathcal{L}_{\text{stft}}$ and $\mathcal{L}_{\text{energy}}$, Flow-HOA effectively flattens the spectral response, ensuring that the timbre of the reconstructed audio remains natural and uncolored.

\subsubsection{Analysis of Spatial Accuracy}
Spatial consistency is paramount for immersion in VR/AR. The baseline exhibits a high {Directional Gain Consistency (DGC)} of 2.17 dB, implying that a sound source would fluctuate unnaturally in volume as it moves around the listener (or as the listener rotates their head). Flow-HOA drastically reduces this variation to {0.84 dB}, achieving a nearly isotropic response despite the irregularity of the microphone array. Furthermore, the improvement in {SPM-KL} (1.14 vs. 1.44) confirms that the spatial energy distribution is sharper. Analytical solutions often suffer from "spatial leakage" where energy spreads into sidelobes; Flow-HOA's spatial loss explicitly penalizes this leakage, resulting in a more focused and accurate acoustic image.

\subsection{Subjective Evaluation}

\subsubsection{Experimental Protocol}
To assess the perceptual validity of the proposed method, we conducted a formal listening test following the MUSHRA (Multiple Stimuli with Hidden Reference and Anchor) paradigm, as defined in ITU-R BS.1534-3. The panel consisted of 16 participants (11 male, 5 female, aged 20--25), all of whom reported normal hearing and had prior experience with critical listening tasks.
The evaluation was performed using the webMUSHRA framework \cite{schoeffler2018webmushra} in a quiet, controlled environment. Stimuli were presented over Sennheiser HD\,600 open-back headphones.
Crucially, the test material consisted of \textit{real recordings} captured by the physical SPMA prototype in the anechoic chamber, with a live human speaker as the source---completely independent from the synthetic FSD50K-based training data. The speaker was positioned at eight azimuths ($0^\circ$ to $315^\circ$ at $45^\circ$ intervals) and recorded separately for each direction. These real four-channel microphone signals were then encoded by both the ASM baseline and Flow-HOA filter banks, and the resulting HOA signals were rendered to binaural audio using the non-individualized Neumann KU\,100 HRTF dataset \cite{bernschutz2013spherical}, with the HRTF forward direction aligned to the $0^\circ$ azimuth of the microphone array coordinate system. No head tracking was employed.
Participants evaluated two attributes on a continuous 0--100 scale: (1) \textit{overall sound quality}, for which they rated the perceived fidelity relative to a hidden reference, with a 3.5\,kHz low-pass filtered anchor as the lower bound; and (2) \textit{spatial localization accuracy}, for which they rated how accurately they could perceive the intended direction of the sound source, with an opposite-angle stimulus as the anchor. A post-hoc screening was applied to exclude listeners who failed to rate the hidden reference above 90, ensuring data reliability.

\begin{figure}[t]
\centering
\includegraphics[width=0.92\linewidth]{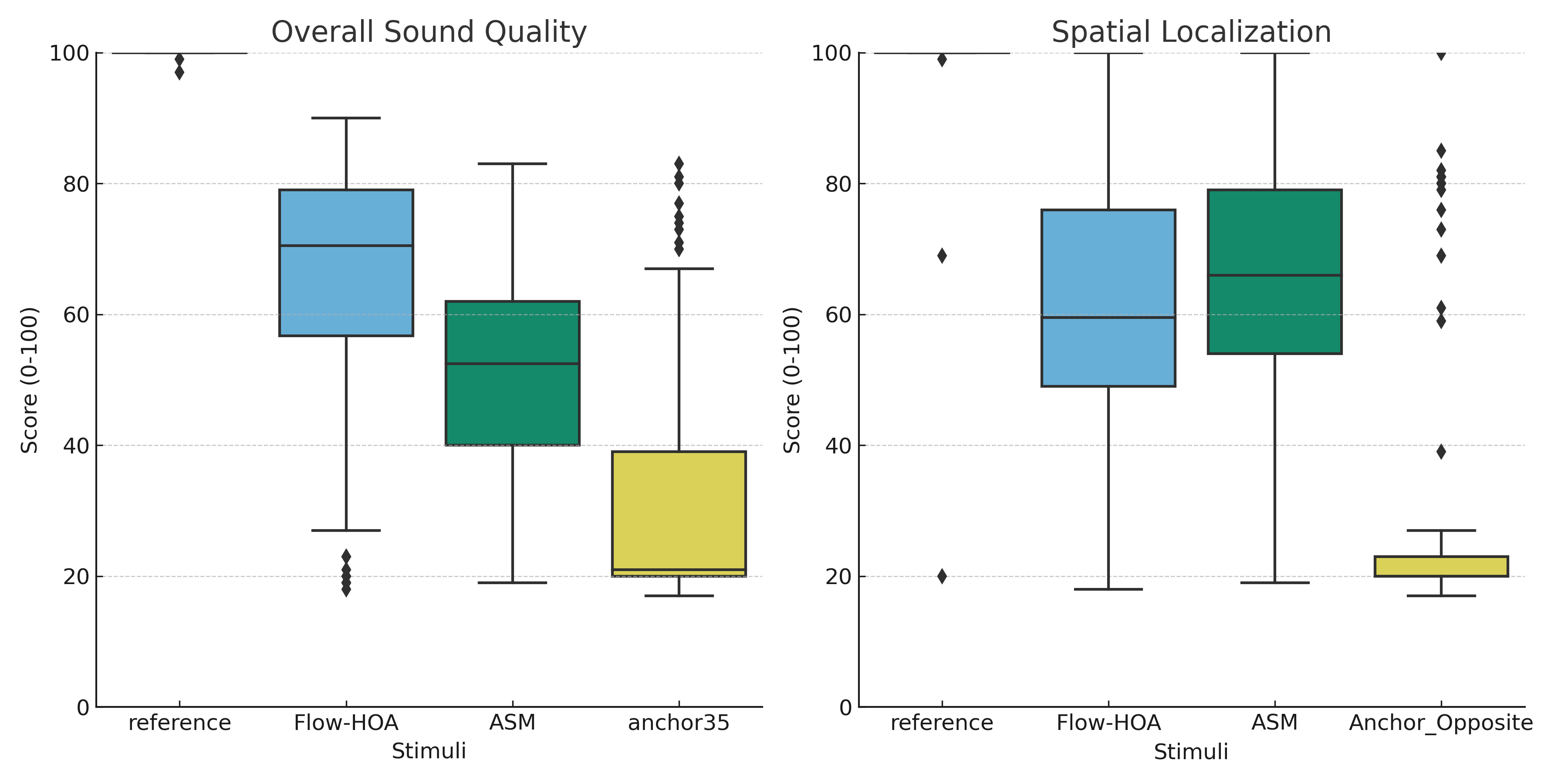}
\caption{Boxplots of the MUSHRA listening test results. The subjective ratings confirm the superiority of Flow-HOA in overall timbre quality, while spatial localization ratings reveal complex psychoacoustic interactions.}
\label{fg:boxplot}
\end{figure}

\subsubsection{Results and Analysis}
The collected ratings (Fig.~\ref{fg:boxplot}) were analyzed using a repeated-measures ANOVA followed by Holm-corrected post-hoc t-tests to determine statistical significance.

\textbf{1) Overall Sound Quality:} The analysis revealed a significant main effect of the processing method ($F(3,45)=145.03, p < .001$). Flow-HOA achieved a mean score of {64.4}, significantly outperforming the ASM baseline (Mean = {50.9}) with a substantial margin of 13.6 points (Holm-corrected $p < .001$). Qualitative feedback from subjects indicated that the ASM outputs often suffered from "metallic" coloration and "phasiness," artifacts typically associated with the comb-filtering effects of analytical beamformers. In contrast, Flow-HOA was consistently rated as "cleaner" and "more natural," corroborating the objective improvements observed in the LSD and SI-SDR metrics.

\textbf{2) Spatial Localization \& The IHL Phenomenon:}
For spatial localization, while a main effect was found ($F(3,45)=194.68, p < .001$), the post-hoc comparison showed {no statistically significant difference} between Flow-HOA (Mean = 60.4) and ASM (Mean = 62.6; $p=.40$).
This counter-intuitive result warrants a deeper psychoacoustic interpretation. Several participants reported that while Flow-HOA provided a highly focused sound image, it sometimes triggered a pronounced {In-Head Localization (IHL)} effect—where the sound is perceived inside the cranium rather than externalized.
This phenomenon is a known limitation when rendering dry, anechoic signals with non-individualized HRTFs \cite{Blauert1997, Hartmann1996}. Without reverberation cues or personalized spectral notches to aid externalization, the brain struggles to project the high-fidelity dry source outwards.
Paradoxically, the ``worse'' signal from the ASM baseline may have partially avoided this issue. We hypothesize that the spectral coloration and temporal smearing inherent in ASM introduced pseudo-environmental cues that inadvertently aided externalization. Importantly, IHL is a well-known limitation of non-individualized HRTF rendering without head tracking and is not specific to our encoding method. This finding suggests that Flow-HOA's high fidelity successfully revealed the intrinsic limitations of the anechoic binaural rendering chain, whereas the baseline masked them with distortion. The interaction between encoding fidelity, non-individualized HRTF spectral cues, and perceived externalization is complex and multifaceted; identifying objective measures that reliably predict this subjective--objective discrepancy remains an open research question for future investigation.

\section{Conclusion}
\label{sec:conclusion}
We introduced Flow-HOA, a framework that formulates the design of FIR filters for HOA encoding as a generative joint optimization task. Unlike conventional approaches based on simplified analytical objectives, our method employs conditional flow matching guided by a composite multi-domain loss to learn the mapping from a prior distribution to optimal filters.
Objective experiments on synthetically simulated data showed that Flow-HOA outperformed the ASM baseline in signal fidelity and spatial accuracy. Subjective listening tests conducted on real microphone array recordings---captured independently from the synthetic training data using a live speaker in the anechoic chamber---confirmed higher overall sound quality and demonstrated the model's ability to generalize from synthetic training conditions to real-world capture. The results also revealed that improved fidelity does not necessarily enhance spatial externalization, as in-head localization was still observed---a known limitation of non-individualized HRTF rendering without head tracking. While the training and objective evaluation rely on synthetically generated data from measured impulse responses convolved with dry audio, the subjective results on real recordings provide initial evidence of practical applicability. Nonetheless, validation on multi-source recordings in reverberant environments, as well as integrating externalization-aware metrics into the optimization process, remains important future work.
\section*{Acknowledgment}
This work is supported by the Shenzhen Science and Technology Program (No.\ ZDCY20250901103533010), and the High-performance Computing Platform of Peking University.

\bibliographystyle{IEEEbib}
\bibliography{refs}

\end{document}